%% file: ms.tex
\documentclass[sigconf]{acmart}

\usepackage{booktabs} % For formal tables
\usepackage{balance}       % to better equalize the last page
\usepackage{balance}       % to better equalize the last page
\usepackage{graphicx}
\usepackage{textcomp}
\usepackage{blindtext}
\usepackage{subfig}
\usepackage{dblfloatfix}

\usepackage{booktabs}
\usepackage{multirow}
\usepackage{siunitx}
\usepackage{microtype}        % Improved Tracking and Kerning
\usepackage[all]{hypcap}    % Fixes bug in hyperref caption linking
\usepackage{ccicons}          % Cite your images correctly!
\usepackage{todonotes}
\usepackage{graphicx}
\usepackage{type1cm}
\usepackage{eso-pic}

\setcopyright{acmlicensed}
\copyrightyear{2019} 
\acmYear{2019} 
\setcopyright{acmlicensed}
\acmConference[SIGCSE '19]{Proceedings of the 50th ACM Technical Symposium on Computer Science Education}{February 27-March 2, 2019}{Minneapolis, MN, USA}
\acmBooktitle{Proceedings of the 50th ACM Technical Symposium on Computer Science Education (SIGCSE '19), February 27-March 2, 2019, Minneapolis, MN, USA}
\acmPrice{15.00}
\acmDOI{10.1145/3287324.3287443}
\acmISBN{978-1-4503-5890-3/19/02}

\begin{document}
\title{Collaboration Versus Cheating}
\subtitle{Reducing Code Plagiarism in an Online MS Computer Science Program}

\author{Tony Mason}
\authornote{Also affiliated with the University of British Columbia (fsgeek@cs.ubc.ca)}
\orcid{0000-0002-0651-5019}
\affiliation{%
  \institution{Georgia Institute of Technology}
}
\email{fsgeek@gatech.edu}

\author{Ada Gavrilovska}
\affiliation{%
\institution{Georgia Institute of Technology}
%\city{Atlanta}
%\state{Georgia}
%\country{US}
}
\email{ada@cc.gatech.edu}

\author{David A. Joyner}
\affiliation{%
  \institution{Georgia Institute of Technology}
}
\email{david.joyner@gatech.edu}

\renewcommand{\shortauthors}{T. Mason et al.}

\begin{abstract}
  We outline how we detected programming plagiarism in an introductory online course for a
  master's of science in computer science program, how we achieved a statistically 
  significant reduction in programming plagiarism by combining a clear explanation of
  university and class policy on academic honesty reinforced with a short but formal
  assessment, and how we evaluated plagiarism rates before
  and after implementing our policy and assessment.
\end{abstract}
  
%
% Generated this code.  Should review it
%



\maketitle

\input{gios-plagiarism}

% \pagebreak

\balance
\bibliographystyle{ACM-Reference-Format}
%\bibliography{sample-bibliography}
\bibliography{plagiarism.bib} 

\end{document}

%% file: gios-plagiarism.tex
%% Note: maximum length is 6 pages + 1 page of references

\section{Introduction}\label{Introduction}

\begin{figure*}[!bt]
 \centering
 \includegraphics[width=1.75\columnwidth]{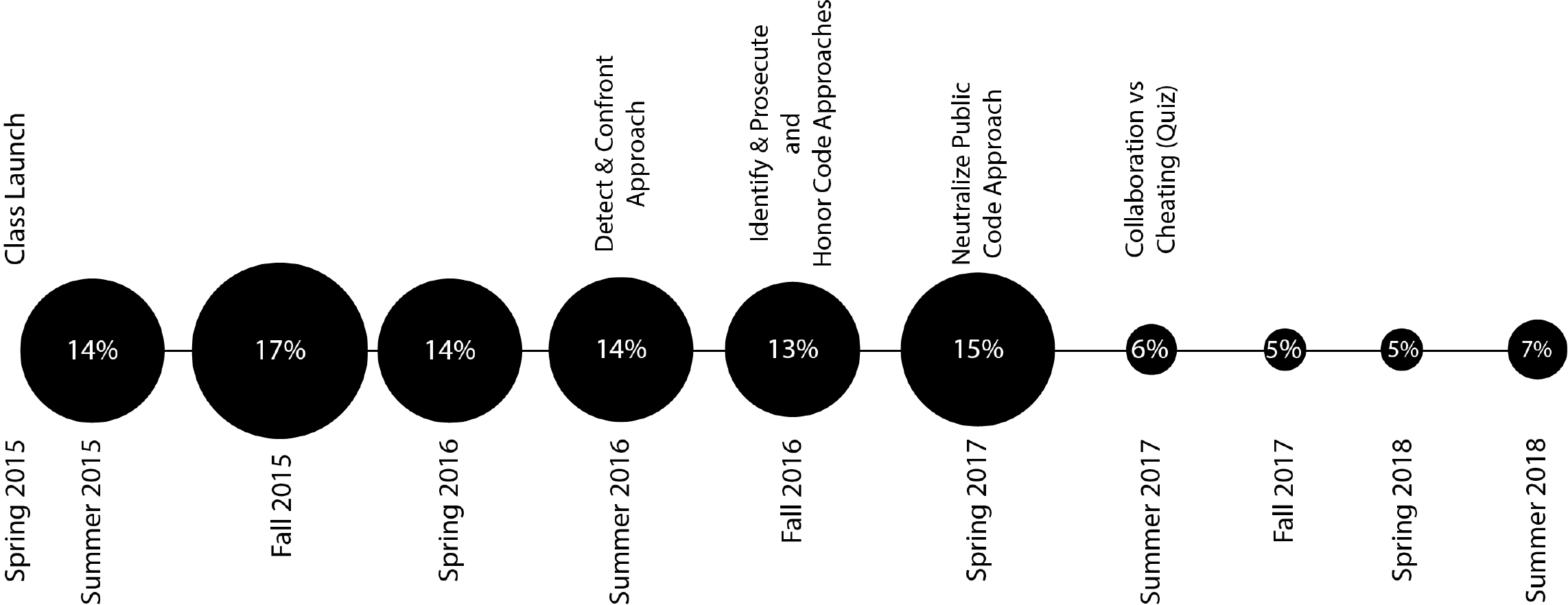}
 \caption{Course timeline}\label{timeline}
 Course timeline and percentage of unique students detected via automatic plagiarism checks. See Table \ref{table:unique-students} for detailed data. 
 
 A student flagged multiple times is only counted \textit{once}.
 \end{figure*}

Academic honesty is critical for those engaged in research. Because we build upon the work of others, we must be able to rely on the integrity of work done by others:

\begin{quotation}
\textit{
As a society, we rely on the academic and journalistic integrity of other people's work. The whole point of academic research is to share knowledge with others and learn from one another. Since knowledge and ideas are the primary product produced by academic communities, it is essential that this knowledge is accurate and gives credit to those who created it~\cite{u-ontario-academic-honesty}}.
\end{quotation}

The concept of intellectual integrity is not restricted to academic work; dishonest behavior spills over into the non-academic world. Cheating is a difficult habit to break, and forms of cheating (such as plagiarism) spread easily when left unchallenged~\cite{crittenden2009cheating}.

While instructors for complex technical classes focus on promoting mastery, students often focus on achieving the highest possible marks. When they have not achieved mastery of the material, they may submit the work of others as their own. The issue of academic honesty in programming classes is not a new problem, having been reported early in the history of computer science education\cite{Grier:1981:TDP:800037.800954}. The problem of plagiarism is complicated by the fact that the division between collaboration and cheating for programming is not clear; modern work practice is often to modify existing code to fit the current need. 

Plagiarism in which students copied the work of those who took the same course in earlier semesters was problematic in our online program, which grants a master of science degree in computer science to students who complete it. The program began in 2014, and all courses and student interactions are conducted online~\cite{goodman2017can}. As of Spring 2018 there were 6,365 students enrolled, with most (70.2\%) being US citizens or permanent residents; 85.1\% were men, and 14.8\% were from under-represented minorities. Most students were enrolled in a single class, in which total course enrollments reached 8,737.\footnote{https://www.omscs.gatech.edu/prospective-students/numbers}

One challenge in delivering computer science classes at this scale is implementing an effective formative analysis of students' comprehension without unduly burdening the instructional team. Some classes in our program create unique projects each term, but this is not practical for our class, as we provide automated feedback via a ``black box'' testing mechanism, which requires considerable effort to modify, let alone rewrite from scratch each semester. Similarly, we found that a strict enforcement mechanism was effective in suppressing the repeated instances of academic dishonesty from the same students, but substantial effort was required to follow the \textit{due process} requirements of the university; we observed that some instructors ignored problems rather than dedicate the considerable effort required to vigorously pursue them.

We were motivated to conduct this research because we observed that plagiarism in our course was an increasing problem. We discussed this situation internally, reviewed techniques for mitigating the issue, and even implemented several suggested mechanisms. While we found that rapid response and enforcement was effective for \textit{subsequent} projects, we sought a mechanism that would discourage plagiarism \textit{before} it became a burden on the instructional team. In addition, fewer cases up front would make it easier to enforce the class policy simply because there would be fewer such cases to report to the university.

In Summer 2017 we introduced a single change: a formal assessment, given via our learning management system (LMS), that spelled out the expected behavior from students in our class, why we viewed it as an important aspect of the course, and the potential ramifications for students that violated the policy. We observed a substantial drop in the number of detected cases. At that point we decided to perform a more rigorous evaluation of our results. This led to a thorough review of our historical data, as well as an ongoing review of subsequent classes, which allowed us to determine whether we were, in fact, observing a valid trend. We now have four semesters of data since we first implemented the formal assessment. We have observed a statistically significant decrease in the rates of plagiarism in the courses based upon our approach of using a quiz.  Figure \ref{timeline} shows our 
measured rate of plagiarism over time and the effect of our change beginning in Summer 2017.  Table \ref{table:honor-code}
demonstrates our claim this represents a statistically significant decrease in plagiarism.

Our approach combines a clear definition of the course expectations and the university's honor code. This is explained in detail to the students in the form of a formal assessment exercise.\footnote{https://github.com/fsgeek/collaboration-versus-cheating} This approach, a low-effort mechanism for clearly communicating expectations in an instructional setting, decreased our objectively measured plagiarism rate. It also represents a novel ``middle ground'' between prior work that showed honor codes were \textbf{not} effective in online classes and one that showed that a comprehensive course in academic honesty, with evaluation, \textit{was} effective. We suggest this approach can be applied successfully to other, similar courses.
  
\section{Background}\label{Background}

\begin{table*}[t]
    % \usepackage{booktabs}
    % \usepackage{multirow}
    \caption{Project 1 Submission \& Plagiarism Data}
    \label{table:one}
    Project 1 submissions consist of four distinct units; each unit is independently analyzed. Per semester, we report the
    total number of submissions and the percentage flagged for plagiarism (as defined in \S \ref{background:plagiarism}).
    \centering
    \resizebox{\textwidth}{!}{%
    \begin{tabular}{@{}rSSSSSSSS@{}}
     \toprule
     \multicolumn{1}{c}{} & \multicolumn{2}{c}{PR1-A} & \multicolumn{2}{c}{PR1-B} & \multicolumn{2}{c}{PR1-C} & \multicolumn{2}{c}{PR1-D} \\ 
     \multicolumn{1}{c}{\multirow{-2}{*}{Semester}} & \multicolumn{1}{c}{Submissions} & \multicolumn{1}{c}{\% Plagiarism} & \multicolumn{1}{c}{Submissions} & \multicolumn{1}{c}{\% Plagiarism} & \multicolumn{1}{c}{Submissions} & \multicolumn{1}{c}{\% Plagiarism} & \multicolumn{1}{c}{Submissions} & \multicolumn{1}{c}{\% Plagiarism} \\ \cmidrule(r){1-9}
     \rowcolor[HTML]{EFEFEF} 
     Summer 2015 & 152 & 3.95 & 145 & 4.14 & 121 & 3.31 & 129 & 5.43 \\
     Fall 2015 & 200 & 9.50 & 194 & 7.73 & 190 & 4.74 & 191 & 6.81 \\
     \rowcolor[HTML]{EFEFEF} 
     Spring 2016 & 174 & 5.75 & 170 & 7.06 & 167 & 3.59 & 162 & 4.94 \\
     Summer 2016 & 150 & 1.33 & 153 & 4.58 & 124 & 4.84 & 123 & 4.07 \\
     \rowcolor[HTML]{EFEFEF} 
     Fall 2016 & 183 & 6.01 & 183 & 6.56 & 156 & 3.85 & 155 & 5.77 \\
     Spring 2017 & 140 & 7.86 & 140 & 10.71 & 119 & 10.08 & 121 & 12.40 \\
     \rowcolor[HTML]{EFEFEF} 
     Summer 2017 & 77 & 1.30 & 70 & 0.00 & 67 & 2.99 & 68 & 0.00 \\
     Fall 2017 & 224 & 2.68 & 225 & 1.78 & 203 & 2.46 & 209 & 1.44 \\
     \rowcolor[HTML]{EFEFEF} 
     Spring 2018 & 274 & 4.01 & 281 & 4.27 & 235 & 1.70 & 245 & 2.45 \\ 
     Summer 2018 & 172 & 3.49 & 171 & 5.26 & 160 & 1.88 & 166 & 1.81 \\ \bottomrule
     \end{tabular}%
    }
   \end{table*}

Our course includes two substantial programming projects, which are generally reused each semester with only minor revisions. While some classes in the program do change their projects from semester to semester, it is not feasible for our class to change our course projects.

Students are given initial ``starter'' code with instructions on the project requirements. They are encouraged to \textit{collaborate} with one another, but to complete and submit their own work. Programs are submitted to a test framework that provides feedback but not final grades. Students are required to submit a \texttt{README} file with the project, explaining their understanding of the project, the challenges they experienced, and \textit{the external sources they used}.

\subsection{Plagiarism}\label{background:plagiarism}

For the purposes of our evaluation, we determined the presence of plagiarism using the results of MOSS\cite{MOSS}, which is a standard 
tool used in plagiarism detection for programming courses in our programs. We invoked MOSS using \texttt{-l c -n 1000 -d -c ``<project 
name>'' -b <template> ... -b <template>}. This set of options tells MOSS that the language (\texttt{-l}) is C\cite{kernighan2006c}, requests the top 1000 ``hits'' (\texttt{-n}), designates files in the same directory (\texttt{-d}) as part of the same program,
adds a unique name to the top of the MOSS output (\texttt{-c}), and ensures that the starter code (\texttt{-b}) is not itself assessed for plagiarism.

In a normal class execution, we manually assess each case of suspected plagiarism. MOSS is a \textit{tool} to identify \textit{potential} cases; results are suggestive of plagiarism, but not definitive. To make a referral to the university, we must be sure that the reported offense is defensible.

In evaluating the effectiveness of our instructional approach for plagiarism reduction, we rejected using the manual evaluation method in order to avoid injecting additional bias into our analysis. Instead, we chose a definitive cut-off of 30\% MOSS similarity. We chose this value because, in reviewing the cases considered by the instructional team, we noted that all instances at or above 30\% had been referred by the instructional staff, no cases below 20\% were referred, and some cases between 20\% and 30\% were referred. We were confident that a demonstrable decrease in the plagiarism rate at this level was strongly supportive of an effective reduction scheme, even if there were a small number of false positive results.

Although our initial choice of 30\% was based on our \textit{ad hoc} analysis of the cases that we had reviewed, prior work suggests this was a reasonable approach, based upon the low likelihood of a false positive over a large number of MOSS tokens\cite{Yan:2018:TUI:3159450.3159490}. The projects in our class typically include several hundred lines of C code written by students. As a result, the total number of unique tokens evaluated by MOSS is quite substantial, and false positives are less likely. 

The original class practice involved evaluating plagiarism using only \textit{intra}-semester data, but we observed that
incorporating inter-semester data revealed 20\% more cases of plagiarism. With this in mind, we converted to using a full cross-semester analysis: we compared all student submissions between Summer 2015 and Summer 2018 (10 semesters). In cases where 
trying to evaluate all submissions over time did not work --- MOSS simply never returned results --- we compared the current semester's submissions against individual prior years' submissions, combining the results at the end for the current semester. We found this result consistent with prior work~\cite{Pierce:2017:ISP:3017680.3017797}.

\subsection{Course Messaging (Pre--Summer 2017)}

The inaugural semester of the course was Spring 2015; there was no prior body of work from which students could draw. Beyond making students aware of the existing university honor code, we did not raise the issue of plagiarism to the class during that semester.

In Spring 2016 we noticed indications of plagiarism and began to investigate. Our first course of action was to augment the syllabus to more clearly explain the course standards for plagiarism. We also offered amnesty to students who came forward to admit their plagiarism, but found amnesty completely ineffective; \textit{none} of the students from the suspected cases came forward. Instead, students who had used small online examples of code came forward.

In Spring 2017 we took aggressive action by quickly completing a review of the code submissions shortly after students completed each project. When we identified specific cases of suspected plagiarism, we scheduled meetings to discuss our findings with the students. We made contact via the online student forum system (Piazza) and conducted online meetings (via WebEx or BlueJeans). Many students ignored our requests to discuss their assignments. Of those who spoke with us, some admitted to plagiarism and some denied it. All cases were submitted to the university under its process for handling the issue. The plagiarism rate between the first and second project plummeted. However, we estimate that each case we prosecuted required more than five hours of instructional team 
time. This approach, while effective, was too labor intensive to scale for a large class.

\subsection{Collaboration versus Cheating}\label{section:collaboration-versus-cheating}

Beginning in Summer 2017, we augmented the existing information about course expectations, the academic honesty policy, and penalties for violating this policy with an additional reinforcement mechanism: a quiz, administered using the university's LMS, that required the students' active acknowledgement that they understood and agreed to the policy. The initial results for Summer 2017 were encouraging: the number of cases of plagiarism noticeably decreased. However, the class size for that semester was the smallest to date. While we were optimistic, we decided it was premature to declare victory.

Subsequent semesters supported our finding that there was a statistically significant decrease in the rate of plagiarism in Summer 2017. We discuss our findings in \S \ref{evaluation} and our analysis in \S \ref{analysis}.

\section{Evaluation}\label{evaluation}

Each project in our course consists of discrete parts: Project 1 has four and Project 2 has two. Each component is individually submitted for grading, and we evaluate each distinct component using MOSS. Table \ref{table:one} reports our results for Project 1 from Summer 2015 through Summer 2018. Table \ref{table:two} reports our results for Project 2 from Summer 2015 through Fall 2017, excluding Spring 2017, as that semester we implemented a different, aggressive-prosecution model. Table \ref{table:aggregate} summarizes our results for all presentations of the course. Table \ref{table:unique-students} summarizes our results for all \textit{unique} students identified as plagiarizing assignments in a given semester.

\subsection{Effectiveness}\label{section:effectiveness}

\begin{table}[]
    \caption{Project 2 Submission \& Plagiarism Data}\label{table:two}
    Project 2 consists of two separate submission units.  Per semester, we report the total number of submissions and the 
    percentage flagged for plagiarism (as defined in \S \ref{background:plagiarism}).   
    \centering
    \resizebox{\columnwidth}{!}{%
    \begin{tabular}{@{}rSSSS@{}}
        \toprule
        \multicolumn{1}{c}{} & \multicolumn{2}{c}{Part 1} & \multicolumn{2}{c}{Part 2} \\ 
        \multicolumn{1}{c}{\multirow{-2}{*}{Semester}} & Submissions & \% Plagiarism & Submissions & \% Plagiarism \\ \cmidrule(r){1-5}
        \rowcolor[HTML]{EFEFEF} 
        Summer 2015 & 136 & 7.35 & 132 & 7.58 \\
        Fall 2015 & 147 & 8.16 & 132 & 8.33 \\
        \rowcolor[HTML]{EFEFEF} 
        Spring 2016 & 142 & 1.41 & 140 & 2.86 \\
        Summer 2016 & 135 & 8.11 & 126 & 7.94 \\
        \rowcolor[HTML]{EFEFEF} 
        Fall 2016 & 155 & 6.45 & 146 & 8.90 \\
        Spring 2017 & 122 & 4.10 & 122 & 3.28 \\
        \rowcolor[HTML]{EFEFEF} 
        Summer 2017 & 71 & 0.00 & 68 & 2.94 \\
        Fall 2017 & 203 & 0.99 & 196 & 0.00 \\
        \rowcolor[HTML]{EFEFEF} 
        Spring 2018 & 251 & 0.00 & 238 & 0.84 \\
        Summer 2018 & 164 & 0.00 & 159 & 3.14 \\
        \bottomrule
    \end{tabular}%
 }
\end{table}

To evaluate our ``Collaboration versus Cheating'' approach, we analyze each separate part of a submitted project by performing a one-tailed two-sample t-test, assuming unequal variances. This approach is appropriate for this evaluation because we are interested only in interventions that yield lower plagiarism rates. We are testing distinct groups of students, but with similar backgrounds (see \S \ref{analysis:demographics} for further discussion). We tested against the null hypothesis with a 95\% confidence (\textit{p < 0.05}).

In evaluating Project 1 cases, we use data from Summer 2015 through Summer 2018. Data from Summer 2015 through Spring 2017 represent the data prior to the introduction of this approach, while those from Summer 2017 through Summer 2018 represent the data since the implementation of this approach. We omit Spring 2017 Project 3 data, as it relates to the strong success of a rapid response scheme, where the instructional team reached out to Project 1 students. No other semester used this approach; it does not surprise us that aggressive enforcement is generally successful.

Our interpretation of these results is that the samples in question are unlikely to come from the same population (\textit{p < 0.05}). We conclude that our intervention in this case was significant. However, we acknowledge the possibility that some other differences account for these changes (this is further discussed in \S \ref{analysis}).

Independently, we examine the number of \textit{unique} students identified per semester. These data are shown in Table \ref{table:unique-students}. The change is clearly visible from the table. We perform a comparable t-test on these results, comparing the Summer 2015 through Spring 2016 results against the Summer 2016 through Summer 2018 results, and once again note that these results are unlikely to come from the same population (\textit{p < 0.05}). 

\subsection{Resource Requirements}\label{section:resource-requirements}

An important consideration for us regarding the role of course instructors is the level of resources required for implementing anti-plagiarism techniques. The time required to add the quiz to the LMS is nominal, and the material can be copied from one semester to another. The post we add to our student engagement forum (Piazza) is also copied from the prior semester. The total effort required for this is less than one hour and is independent of the number of students in the class. Responses to follow-up questions from students and clarification of the policy require approximately an additional hour of time. Fewer than a dozen students asked follow-up questions on average; most questions consisted only of one or two paragraphs and were asked via Piazza\cite{piazza} so that the answers were visible to the entire class.

The time required for the students to complete the evaluation was generally less than 15 minutes. 

In addition, our work for evaluating the plagiarism activity has led us to automate much of the process involved. This is done using a set of scripts that invoke MOSS with all of the current and prior semesters' code as well as any starter code provided to the students. The scripts then collect the results from MOSS, download and ``scrape'' the HTML pages that MOSS provides, and generate output data files in a format that is amenable to further processing. This step can take up to three hours for a class with approximately 300 student submissions. One interesting complication is that one of the project parts is now so large that we must break up the submission process incrementally and submit the current semester submissions against prior submissions grouped by year --- otherwise, MOSS fails to provide results --- a frustrating type of error.

\begin{table}[t]
    \caption{Aggregate Plagiarism Results}\label{table:aggregate}
    MOSS-reported similarity across all semesters: breakdown by project/sub-project, reporting number of flagged submissions, percentage (of total), plus average similarity \% and average line count similarity (with standard deviation).

    \centering
    \resizebox{\columnwidth}{!}{%
    %\begin{tabular}{@{}||c|c|c|S|S|S|S|S||@{}}
    \begin{tabular}{@{}cccSSSSS@{}}
        \toprule
        & \multicolumn{1}{c}{} & \multicolumn{2}{c}{Plagiarism} & \multicolumn{4}{c}{MOSS} \\ \cmidrule(l){3-8} 
        \multirow{-2}{*}{Project} & \multicolumn{1}{c}{\multirow{-2}{*}{Total Submissions}} & \multicolumn{1}{l}{Detected} & \multicolumn{1}{l}{Percent} & \multicolumn{1}{c}{\begin{tabular}[c]{@{}c@{}}Average \\ \%\end{tabular}} & \multicolumn{1}{c}{\begin{tabular}[c]{@{}c@{}}Standard\\ Deviation\end{tabular}} & \multicolumn{1}{c}{\begin{tabular}[c]{@{}c@{}}Average\\ Lines\end{tabular}} & \multicolumn{1}{c}{\begin{tabular}[c]{@{}c@{}}Standard\\ Deviation\end{tabular}} \\ \midrule
        PR1-A & 1647 & 83 & 5.04 & 13.01 & 15.47 & 44.33 & 50.30 \\ 
        \rowcolor[HTML]{EFEFEF} 
        PR1-B & 1623 & 92 & 5.67 & 13.11 & 16.96 & 31.19 & 35.23 \\ 
        PR1-C & 1409 & 57 & 4.05 & 15.77 & 11.51 & 38.04 & 36.75 \\ 
        \rowcolor[HTML]{EFEFEF} 
        PR1-D & 1475 & 69 & 4.68 & 18.10 & 15.01 & 41.81 & 59.94 \\ 
        PR2-A & 1509 & 52 & 3.45 & 10.76 & 11.16 & 30.82 & 25.00 \\ 
        \rowcolor[HTML]{EFEFEF} 
        PR2-B & 1453 & 61 & 4.20 & 8.21 & 15.41 & 62.77 & 100.14 \\ 
        \bottomrule
    \end{tabular}%
    }
\end{table}

Once a case is identified, we review each of the suspect cases to eliminate those that do not appear to be genuine acts of plagiarism. This involves a review of the code and the student's \texttt{README} file. If these indicate a likely case of plagiarism, either a meeting is scheduled with the student --- if this is the first time we have observed an issue for this student --- or we refer
the student to the university. Since the entire program is online, student meetings in this context are via an online video conferencing service. At least two members of the instructional team are present and, with the student's permission, the meeting is recorded. The information from MOSS is presented to them, and we discuss the similarity. If they admit to plagiarism, we give them a grade of zero points on the suspected sections of the project. The outcome is reported to the university, which tracks the issue so that repeated offenses can be handled appropriately while safeguarding the students' privacy rights.

In cases where students ignore our request for a meeting (which frequently happens), tell us they have withdrawn from the course, or do not admit to the plagiarism, we process a formal referral to the university. The referral includes a copy of the plagiarism quiz, the syllabus explaining the course policy, and the code identified by MOSS. Once a decision is made, the instructional team is notified. If a grade change is required, a paper form must be completed and signed by various members of the department. Overall, we estimate that prosecuting a case requires more than five hours of instructional team time.

Even the most superficial analysis indicates that any approach that minimizes plagiarism is an improvement over the simple enforcement model. Thus, we defer an actual cost calculation for this improvement.

\section{Analysis}\label{analysis}

\begin{table}[]
    \caption{``Collaboration versus Cheating'' Assessment T-Test Results }\label{table:honor-code}
    See \S \ref{section:collaboration-versus-cheating}. One-tailed two-sample t-test, assuming unequal variances: demonstrates statistical significance of intervention with respect to pre-intervention classes.

    \centering
    \begin{tabular}{@{}rSSS@{}}
    \toprule
    \multicolumn{1}{l}{Project} & \multicolumn{1}{c}\textit{T}-crit & \multicolumn{1}{c}\textit{T}-stat & \multicolumn{1}{c}\textit{P(\textbf{$\leq$0.05})} \\ \midrule
    PR1-A & 2.23 & 0.03 & \textbf{0.032} \\
    \rowcolor[HTML]{EFEFEF} 
    PR1-B & 2.90 & 1.89 & \textbf{0.011} \\
    PR1-C & 2.83 & 1.94 & \textbf{0.015} \\
    \rowcolor[HTML]{EFEFEF} 
    PR1-D & 3.93 & 1.94 & \textbf{0.004} \\
    PR2-A & 4.70 & 2.13 & \textbf{0.002} \\
    \rowcolor[HTML]{EFEFEF} 
    PR2-B & 4.09 & 1.89 & \textbf{0.004} \\
    \bottomrule
    \end{tabular}%
\end{table}

We are aware of several open questions that relate directly to the veracity of our results.

\subsection{Hidden Plagiarism}\label{analysis:hidden_plagiarism}

One possible explanation for the observed decrease in the plagiarism rate is that students became better at hiding their plagiarism. In an attempt to look for more subtle forms of plagiarism, we have supplemented our work with MOSS by adding a number of other mechanisms, including code watermarks and SHA-2 checksums, which detect tampering with the code testing environment. We have investigated several suspicious cases, but we found no new cases of plagiarism using these mechanisms. A student brilliant enough to spoof SHA-2 checksums is unlikely to need to plagiarize the code for an introductory graduate operating systems course~\cite{rasjid2017review}.

In the unlikely case that students develop the level of insight necessary to circumvent the various plagiarism checks in the system, they must do so independently of other students, since MOSS detects collaborative plagiarism. Thus, to circumvent MOSS, a student must find a candidate code base and then substantially alter its structure to avoid detection. Even if a student were to actively use MOSS to confirm that their code no longer appeared structurally similar to the original code they plagiarized, they would then need to also make sure that their modified code \textit{worked}, which in turn means fixing it and debugging problems that arose due to the substantive changes. Our pedagogical goal is to ensure that students understand the underlying mechanisms; this goal is achieved if the student must understand their now modified code base sufficiently well to pass the tests of their code. This effect has been separately noted, particularly with respect to MOSS and programming classes~\cite{genchang1234_2015}.

One intriguing approach to detect such cases is to perform a more substantive analysis of the student submissions. Our testing mechanism saves all student submissions, which would permit us to evaluate the \textit{history} of student code submissions more thoroughly using existing techniques~\cite{Yan:2018:TUI:3159450.3159490}.

\subsection{Demographic Shifts}\label{analysis:demographics}

\begin{table}[]
    \caption{Unique number of students detected per semester}\label{table:unique-students}
    \centering
    \begin{tabular}{lSS}
        \toprule
        Summer 2015	& 21 & 13.8 \\
        \rowcolor[HTML]{EFEFEF}
        Fall 2015 & 27 & 16.8 \\
        Spring 2016 & 21 & 14.2 \\
        \rowcolor[HTML]{EFEFEF}
        Summer 2016	& 20 & 14.1 \\
        Fall 2016 & 23 & 13.1 \\
        \rowcolor[HTML]{EFEFEF}
        Spring 2017 & 20 & 14.6 \\
        Summer 2017	& 4 & 5.6 \\
        \rowcolor[HTML]{EFEFEF}
        Fall 2017 & 11 & 5.1 \\
        Spring 2018 & 15 & 5.9 \\
        \rowcolor[HTML]{EFEFEF}
        Summer 2018 & 12 & 7.0 \\
        \bottomrule
    \end{tabular}%
\end{table}

One insightful reviewer of our research suggested that our results might be due to shifting demographics. We do not have demographics for the class, but we do have them for the program (see \S \ref{Introduction}). While there has been a small change in demographics, it seems unlikely that the effectiveness of our approach is due to a modest (10\%) change in program demographics.

\subsection{Work for Hire}\label{analysis:work_for_hire}

A reviewer of our research indicated we did not address work-for-hire-style cheating solutions. We agree that this is a weakness of our work. We are encouraged by work being conducted by our colleagues within the program to build tools for identifying such code. It also occurs to us that their work might be combined with the analysis-over-time approach suggested in \S \ref{analysis:hidden_plagiarism} on the theory that a student who contracts out that work would likely only submit it a few times on their own. We can also track the IP address used to submit student work to detect cases where the work-for-hire is given access to the student's account for the purposes of submitting their work to our automatic grading system.

However, we argue that our own work is a valuable tool for decreasing the rate of plagiarism, even if it does not completely eliminate plagiarism.

\subsection{Strict Enforcement}\label{analysis:strict}

One possible alternative explanation for the efficacy of our approach is the more rigorous enforcement model.  While
we have not formally proven this is not a factor, from Summer 2017 through Summer 2018 our enforcement was not as
rigorous as it was in Spring 2017.  Since Summer 2018 we have made significant strides in automating the process
of generating detailed descriptions of our findings, which has made contacting students and preparing referrals to
the university less time-consuming.  Our scripts anonymize the MOSS output by replacing student names with unique
identifiers, convert the HTML pages to PDF documents, and construct one directory per current-semester student with
all the relevant information. Our hope is this will further reduce our plagiarism rate.

\section{Prior Work}\label{prior-work}

The basic concepts around plagiarism are hardly new and have been extensively reviewed~\cite{howard1995plagiarisms, howard1999standing, maurer2006plagiarism}. Indeed, the literature contains clear discussions on this problem \textit{with respect to programming courses} decades ago~\cite{Grier:1981:TDP:800037.800954}. While we still do not fully understand online-instruction dynamics, the available evidence suggests that plagiarism occurs in both online and in-person programs~\cite{boyatt2014use,shapiro2017understanding, grijalva2006academic,zhang2013benefiting,yair2014saw,CrimsonCheating2017,daly2005patterns,krieger_2016}. While much of the prior work focused on \textit{undergraduate} academic dishonesty, we did not find any prior work suggesting that these findings do not apply to course-based graduate programs as well.

Similarly, the idea of clarifying the expectations for students in programming classes is not new~\cite{Gibson09softwarereuse,Simon:2018:ISA:3160489.3160502}. Our research augments this prior work by evaluating these techniques in an actual classroom setting over a period of 10 semesters.  

Motivating students to adhere to academic honesty policies and the instructional strategies that effectively achieve this goal, particularly with respect to plagiarism in computer science classes, remains an open area of research and practice~\cite{dagli2017relationships}. Although some previous studies have explored the root-cause analysis for plagiarism, our goal is not to understand \textit{why} students plagiarize, but rather to understand how to discourage it through a combination of detection and education mechanisms~\cite{eaton2017plagiarism}. Our approach is consistent with prior studies suggesting a move away from a moralistic view of plagiarism, instead reframing discussion on plagiarism as an integral part of the learning process. In other words, \textit{even if} students do not understand what plagiarism is when starting the program, educating them about it is important and beneficial, albeit a secondary goal to the instructors'~\cite{adam2017s}. That said, our research also supports the prior observation that simply informing students about an honor code \textit{does not} lead to a decrease in plagiarism in online programs~\cite{mastin2009online}. Active education on academic honesty through formal training, however, \textit{may be} effective~\cite{curtis2013online}.

We admit that we do not know the optimal mechanism for motivating students to adhere to academic honesty policies. We have not explored the specific issue of motivations behind plagiarism (which the literature suggests can be cultural, economic, or perception-based), but we do note that our successful interventions are consistent with educating students, regardless of their cultural background~\cite{cosma2017perceptual, baker2017ignorance, chien2017taiwanese}.

There is a strong body of research describing techniques for reducing plagiarism, but there is a much smaller pool of evaluations regarding the application of these techniques~\cite{macfarlane2014academic,ewing2016addressing,smedley2015intervention}. One side-effect of this paucity of research is that the most common strategy for combating plagiarism is to shift responsibility to individual instructors, which results in haphazard enforcement and poor knowledge-sharing. By providing a rigorous review of techniques applied over time to the same class, with similar student populations, we contribute solid evidence of techniques that work. We hope that universities can incorporate this evidence into a broader initiative toward improving \textit{all} courses, rather than relying on anecdotal knowledge shared haphazardly between individual instructors~\cite{miss2017development,singh2017effectiveness}.

\section{Future Work}\label{section:future-work}

Besides the obvious possibilities raised during our analysis (see \S \ref{analysis}), we can see several areas of future work:

\begin{itemize}
    \item We can augment the mechanisms available for detecting plagiarism. While MOSS is an excellent tool, it is not necessarily applicable to all programming classes. Evaluating new mechanisms for detecting code plagiarism will be useful and is an active area of research~\cite{caliskan2015anonymizing,caliskan2015coding,yang2017authorship,meng2016fine,meng2017identifying,jamil2017automated,duan2017plagiarism,delev2017comparison,victor2016code,kikuchi2015source,heres2017source}. 
    \item We can work on identifying new ways to modify the behavior of students to achieve our goals of further improving academic honesty. 
    \item We can improve our own analysis of the output using our existing tools. Our work on this project has already encouraged us to develop several tools we can use to ease the evaluation of student code in the future, and we suspect there is further work to be done in this area, particularly with an eye toward simplifying the process for others. Similar augmentations over MOSS have been implemented by others as well~\cite{sheahen2016taps,Yan:2018:TUI:3159450.3159490}
    \item We could explore \textit{why} this approach works. We speculate that it is because of the context in which students are presented with the information --- it is presented as a formal quiz, which underscores its importance to students, whereas posts in Piazza are more likely to be viewed as background noise. Studying different ways of presenting this information to the class and measuring the outcome might be possible, but it may also raise ethical research considerations.
\end{itemize}

We also realize that future classes may learn of the techniques that the course instructors use to detect plagiarism. As a result, course instructors will need to find ways to further improve their detection abilities. At some point, the work required to circumvent these checks becomes greater than the effort of simply doing the assignment --- and hopefully achieves the instructors' pedagogical goal~\cite{genchang1234_2015}.

\section{Conclusion}

Prior work demonstrates that honor codes alone are not an effective mechanism for reducing code plagiarism; this research demonstrates that explaining and reinforcing lessons in academic honesty results in statistically significant (\textit{p < 0.05}) decreases in plagiarism rates across all of the distinctive programming submissions in a large online graduate computer science course. In addition, this technique requires minimal effort from the instructors --- in contrast to a strict enforcement policy, which substantially increases the burden on the instructional staff.

\nocite{joyner2016unexpected}
\nocite{sheahen2016taps}
\nocite{macfarlane2014academic}
\nocite{StackOverflow}
\nocite{GitHubDMCA}
\nocite{CrimsonCheating2017}
\nocite{tabsh2015past}
\nocite{yair2014saw}
\nocite{corrigan2015deterring}
\nocite{singh2017effectiveness}
\nocite{corrigan2015deterring}
\nocite{grijalva2006academic}
\nocite{crittenden2009cheating}
\nocite{mastin2009online}
\nocite{curtis2013online}
\nocite{rasjid2017review}
\nocite{corrigan2015measuring}
\nocite{caliskan2015coding}
\nocite{caliskan2015anonymizing}
\nocite{yang2017authorship}
\nocite{meng2017identifying}
\nocite{meng2016fine}

% ICER feedback below
\nocite{cook2014student}
\nocite{Simon:2018:ISA:3160489.3160502}
\nocite{sheard2016negotiating}
\nocite{maurer2006plagiarism}
\nocite{howard1995plagiarisms}
\nocite{howard1999standing}

% SIGCSE references
\nocite{Pierce:2017:ISP:3017680.3017797}
\nocite{Grier:1981:TDP:800037.800954}
\nocite{Gibson09softwarereuse}
\nocite{Yan:2018:TUI:3159450.3159490}